\documentclass[final,twocolumn]{elsarticle}

\PassOptionsToPackage{breaklinks}{hyperref}
\usepackage{hyperref}
\usepackage{amsmath}

\graphicspath{{figures/}}

\begin{document}

\begin{frontmatter}

\title{Monte Carlo calculation of the average neutron depolarization for the NPDGamma experiment}
\author[utkaddress,ornladdress]{K. B. Grammer\corref{correspondingauthor}}
\cortext[correspondingauthor]{Corresponding author. Oak Ridge National Laboratory, PO BOX 2008 MS6466, Oak Ridge, TN 37831-6466}
\ead{grammerkb@ornl.gov}

\author[ornladdress]{J. D. Bowman}

\address[utkaddress]{University of Tennessee, Knoxville, TN, USA}
\address[ornladdress]{Oak Ridge National Laboratory, Oak Ridge, TN, USA}

\begin{abstract}
The NPDGamma experiment measures the asymmetry in $\gamma$-ray emission in the capture of polarized neutrons on liquid parahydrogen. The beam polarization is measured using $^3$He spin analysis, but this measurement does not account for the contribution of depolarization from spin-flip scattering primarily due to orthohydrogen in the bulk liquid. This is a systematic effect that dilutes the experimental asymmetry and is modeled using Monte Carlo. Methods for tracking neutron spin in MCNPX were developed in order to calculate the average neutron polarization upon capture for use as a multiplicative correction to the measured beam polarization for the NPDGamma experiment.\tnotetext[mytitlenote]{This manuscript has been authored by UT-Battelle, LLC under Contract No. DE-AC05-00OR22725 with the U.S. Department of Energy. The United States Government retains and the publisher, by accepting the article for publication, acknowledges that the United States Government retains a non-exclusive, paid-up, irrevocable, worldwide license to publish or reproduce the published form of this manuscript, or allow others to do so, for United States Government purposes. The Department of Energy will provide public access to these results of federally sponsored research in accordance with the DOE Public Access Plan (http://energy.gov/downloads/doe-public-access-plan).}\tnotetext[titlenote2]{\textsuperscript{\textcopyright} 2019. This manuscript version is made available under the CC-BY-NC-ND 4.0 license \href{http://creativecommons.org/licenses/by-nc-nd/4.0/}{http://creativecommons.org/licenses/by-nc-nd/4.0/}}
\end{abstract}


\end{frontmatter}


\section{Introduction}
The NPDGamma experiment~\cite{Blyth2018} uses a polarized neutron beam in order to measure the parity violating asymmetry in the angular distribution of the emitted 2.2~MeV $\gamma$-rays from neutron capture on liquid parahydrogen~\cite{Gericke2011}
\begin{equation}
\vec{\textrm{n}}+\textrm{p}\to\textrm{d}+\gamma.
\end{equation}
The physics asymmetry, $A_\gamma$, is defined by
\begin{equation}
\label{eq:npdg_diffxs}
\frac{d\sigma}{d\Omega} \propto \frac{1}{4 \pi}(1+A_\gamma \hat{\sigma}_{\textrm{n}} \cdot \hat{k}_\gamma).
\end{equation}
where $\hat{\sigma}_{\textrm{n}}$ is the neutron spin, and $\hat{k}_\gamma$ is the $\gamma$-ray momentum.
Neutrons exit the $10\times12$~cm neutron guide on the Fundamental Neutron Physics Beamline (FNPB)~\cite{Fomin2015} and enter a supermirror polarizer~\cite{Balascuta2012}, after which is the NPDGamma apparatus~\cite{Gericke2006, Fry2017}. Polarized neutrons then enter the resonant frequency spin rotator (RFSR)~\cite{Seo2008} and are rotated according to a pattern ($\uparrow\downarrow\downarrow\uparrow\downarrow\uparrow\uparrow\downarrow$) that cancels beam power fluctuations to second order and the signals from opposite spin states are used to isolate the asymmetry signal. The neutrons are incident on the 16-liter liquid hydrogen target~\cite{Santra2010} that is surrounded by 48 CsI(Tl) detectors arranged in 4 rings of 12 detectors each~\cite{Gericke2005a} that detect the $\gamma$-rays from neutron capture. The average neutron beam polarization is measured with an empty liquid hydrogen target vessel, and a $^3$He spin filter~\cite{Musgrave2018} followed by a $^3$He transmission monitor~\cite{Szymanski1994} downstream of the vessel are used to measure the beam polarization. A uniform 9.5 gauss magnetic field aligned within 3 mrad to the vertical axis ($\hat{z}$) transports polarized neutrons to the target vessel~\cite{Blyth2018}.

The raw asymmetry for a given spin sequence is determined using the super ratio method for each pair of detectors, $i$, where a pair of opposed detectors, $l=0\ldots5$ and $m=l+6$, is separated by $180^\circ$, 
\begin{equation}
A_{\textrm{raw}}^{i} = \frac{\sqrt{\alpha_i}-1}{\sqrt{\alpha_i}+1},
\end{equation}
where $\alpha_i$ is the super ratio given by
\begin{equation}
\alpha_i = \frac{N_{\uparrow}^{l}}{N_{\downarrow}^{l}} \frac{N_{\downarrow}^{m}}{N_{\uparrow}^{m}},
\end{equation}
and $N_{\uparrow}$ and $N_{\downarrow}$ are the spin up and spin down signals, respectively.
The physics asymmetry~\cite{Gericke2011} for a given detector is given by
\begin{equation}
A_\gamma = \frac{A_{\textrm{raw}}^{i} - A_{\textrm{app}}^{i}}{P_{\textrm{n}} \Delta^{i}_{\textrm{dep}}(\lambda) \Delta_{\textrm{sf}}},
\label{eq:npdg_physasym}
\end{equation}
where $A_{\textrm{app}}^{i}$ is a term encompassing apparatus asymmetries including noise and beam fluctuations, $P_{\textrm{n}}$ is the measured neutron beam polarization, $\Delta_{\textrm{sf}}$ is the spin flip efficiency, and $\Delta^{i}_{\textrm{dep}}(\lambda)$ is the detector-dependent neutron depolarization correction. The depolarization correction depends on the geometry of the target, detector array, and neutron beam and on the neutron wavelength. Since the detector rings are azimuthally symmetric about the target, the polarization correction can be expressed as a function of detector rings ($\Delta^{r}_{\textrm{dep}}(\lambda)$) rather than a function of detectors or detector pairs.

There are two spin isomers of the hydrogen molecule: parahydrogen $(J=0,2,4\ldots)$ and orthohydrogen $(J=1,3,5\ldots)$. Room temperature hydrogen gas has approximately a 3:1 ortho-to-para ratio, with the equilibrium parahydrogen concentration increasing as the temperature decreases~\cite{Dennison1927}. The energy separation between the $J=0$ ground state of parahydrogen and the $J=1$ lowest energy orthohydrogen state is 14.7~meV and higher order rotational energy levels in the hydrogen molecule are given by
\begin{equation}
E_{\textrm{rot}}(J) = C J(J+1),
\end{equation}
where $C=7.36\textrm{~meV}$ is the rotational constant for the hydrogen molecule rigid rotor and $J$ is the angular momentum of the hydrogen molecule. The energy separation between the lowest rotational energy levels in the hydrogen molecule is used as the criterion for discerning scattering event types in MCNPX. The hydrogen target vessel is initially filled with normal hydrogen gas. The hydrogen gas is then liquefied and operated at 15.6~K and the slow natural conversion process is accelerated by circulating the liquid through an ortho-para converter (OPC)~\cite{Barron-Palos2011} containing 150 ml of hydrous iron (III) oxide 30–50 mesh powder\cite{catalyst} as a catalyst. Production data is taken after the hydrogen in the vessel has been condensed and circulated through the OPC over the course of weeks such that the ortho-para ratio has reached a steady state and is assumed to be constant throughout the bulk liquid.

\begin{figure}[h!] 
\includegraphics{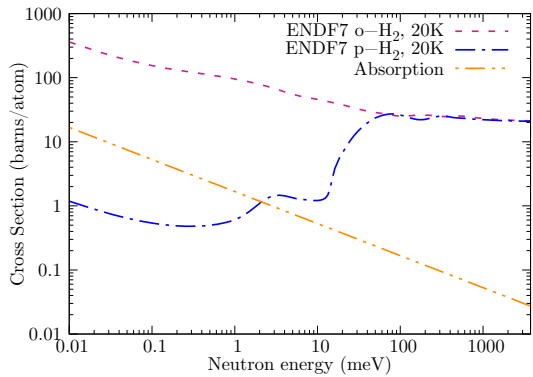}
\caption{Neutron scattering cross sections on parahydrogen and orthohydrogen at 20~K from ENDF-VII~\citep{Chadwick2011} and the neutron absorption cross section~\citep{mughabghab2006atlas}.}
\label{fig:hydrogen_cross_sections}
\end{figure}

The orthohydrogen scattering cross section for neutrons with energy 1-10~meV (figure \ref{fig:hydrogen_cross_sections}) is approximately 2 orders of magnitude larger than both the parahydrogen scattering cross section~\cite{Chadwick2011} and the hydrogen absorption cross section~\cite{mughabghab2006atlas}. The transmission of neutrons through the target as well as the average polarization of neutrons as they propagate through the target is strongly sensitive to the orthohydrogen concentration of the bulk hydrogen. Neutrons can undergo spin flip scattering in the apparatus components and from scattering on orthohydrogen in the bulk liquid hydrogen. Since there is no reliable means of measuring the spin flip scattering rate, the depolarization effect is modeled using Monte Carlo.

There was no in situ method to determine the relative orthohydrogen to parahydrogen concentration for the NPDGamma experiment. The methods to monitor the orthohydrogen concentration available at the time of the construction of the apparatus required feeding electrical signals out of the cryogenic region and hydrogen safety requirements limited the number of wires in the vessel. However, conversion towards a steady state condition with a time constant on the order of days is observed via measurement of the relative neutron transmission and produced a measurement of the parahydrogen scattering cross section~\cite{Grammer2015}. The cross section measurement also yielded a determination of an upper bound (0.15\%) on the orthohydrogen concentration in the bulk liquid while the lower limit (0.015\%) is given by thermodynamic equilibrium at 15.6~K. The upper and lower bounds on the orthohydrogen concentration are used to as inputs for the calculation of $\Delta^{r}_{\textrm{dep}}$.

The polarization correction, $\Delta^{r}_{\textrm{dep}}$ in equation \ref{eq:npdg_physasym}, is calculated with two methods: the first uses a modified version of MCNPX 2.7~\cite{Pelowitz2011}  and the second uses the Young and Koppel free gas model~\cite{Young1964}. These two models will be described next and the two calculation methods are combined to determine the geometry and wavelength dependent polarization for the NPDGamma analysis.

\section{Spin-flip scattering formalism}
We follow the spin-flip scattering formalism outlined in Moon et al.~\cite{Moon1969a}. The double differential cross section for a scattering process with momentum transfer $\vec{K} = \vec{k} - \vec{k}'$ that changes the neutron spin from $s$ to $s'$ and the scattering system state from $q$ to $q'$ is given by~\cite{Moon1969a}
\begin{equation}
\begin{array}{rcl}
\frac{d^2\sigma^{ss'}}{d\Omega'dE'} & = & \sum_q P_q \sum_{q'} \frac{k'}{k} \|\langle q'| \sum_i e^{\imath \vec{K}\cdot\vec{r_i}}U_{i}^{ss'}|q\rangle\|^2 \\
 & & \times \delta(\Delta E_{\textrm{n}} + \Delta E_q), \\
\end{array}
\end{equation}
where $P_q$ is the probability the system is in state $q$, with a sum over the atomic sites $i$, and the $\delta$ function ensures energy conservation for the change in energy of the neutron ($\Delta E_{\textrm{n}}$) and the scattering system ($\Delta E_q$). The scattering amplitude is given by~\cite{Moon1969a},
\begin{equation}
U_{i}^{ss'} = \langle s' | a_{\textrm{coh}, i}-p_i\vec{S_{\perp i}}\cdot\vec{\sigma}+a_{\textrm{inc}, i}\vec{I_i}\cdot{\sigma}| s \rangle,
\label{eq:potential}
\end{equation}
which has a coherent nuclear scattering term ($a_{\textrm{coh}, i}$), a magnetic scattering term ($p_i\vec{S_{\perp i}}\cdot\vec{\sigma}$), and an incoherent nuclear scattering term ($a_{\textrm{inc}, i}\vec{I_i}\cdot{\sigma})$. There are then four combinations of $s$ and $s'$ for neutrons with spin in the $\hat{z}$ direction~\cite{Moon1969a},
\begin{equation}
\begin{array}{rcl}
U^{++} & = & a_{\textrm{coh}, i}-p_i S_{\perp z i}+ a_{\textrm{inc}, i} I_{zi}, \\
U^{--} & = & a_{\textrm{coh}, i}+p_i S_{\perp z i}- a_{\textrm{inc}, i} I_{zi}, \\
U^{+-} & = & -p_i (S_{\perp x i}+\imath S_{\perp y i})+ a_{\textrm{inc}, i}(I_{\perp x i}+\imath I_{\perp y i}), \\
U^{-+} & = & -p_i (S_{\perp x i}-\imath S_{\perp y i})+ a_{\textrm{inc}, i} (I_{\perp x i}-\imath I_{\perp y i}), \\
\end{array}
\label{eq:potentials}
\end{equation}
where $a_{\textrm{coh}}$ is the bound coherent nuclear scattering length, $p_i$ is the magnetic scattering amplitude, $\vec{S}_{\perp}$ is the projection the neutron spin onto the plane perpendicular to the scattering vector, $\vec{I}$ is the nuclear spin, and $a_{\textrm{inc}}$ is the bound incoherent scattering length.

As described by Moon\cite{Moon1969a}, coherent scattering and random isotopic ordering does not contribute to spin-flip scattering. While magnetic scattering contributes to spin-flip scattering in general, it is not relevant for the nonmagnetic hydrogen target in the NPDGamma experiment. Thus, only the incoherent scattering term contributes to spin-flip scattering. The differential scattering cross section per atom becomes simple for the spin-flip and non-spin-flip cases,
\begin{equation}
\frac{d\sigma^{+-}}{d\Omega} = \frac{d\sigma^{-+}}{d\Omega} = \frac{2}{3}a_{\textrm{inc}}^2I(I+1),
\label{eq:sig_sf}
\end{equation}
\begin{equation}
\frac{d\sigma^{++}}{d\Omega} = \frac{d\sigma^{--}}{d\Omega} = \frac{1}{3}a_{\textrm{inc}}^2I(I+1),
\label{eq:sig_nsf}
\end{equation}
and which should be compared to equation \ref{eq:YKxs} below. The spin flip terms for a neutron with spin along the $\hat{z}$ axis, the vertical polarization axis in the NPDGamma experiment, contain only $S_{\perp x i}$ and $S_{\perp y i}$ terms, such that only nuclear spins perpendicular to the neutron spin can cause a spin flip and this gives rise to the $\frac{2}{3}$ factor in equation \ref{eq:sig_sf}. The probability that any given particle interaction leads to a spin flip is given by 
\begin{equation}
P_{\textrm{flip}} = \frac{2}{3}\frac{\sigma_{\textrm{incoh}}}{\sigma_{\textrm{incoh}}+\sigma_{\textrm{coh}}+\sigma_{\textrm{abs}}},
\label{eq:spin_flip_prob}
\end{equation}
where $\sigma_{\textrm{incoh}}$ is the incoherent scattering cross section, $\sigma_{\textrm{coh}}$ is the coherent scattering cross section, and $\sigma_{\textrm{abs}}$ is the absorption cross section.

\subsection{Neutron-hydrogen spin-flip scattering}
\begin{table}
\begin{center}
    \begin{tabular}{| l | c |}
    \hline
    Transition & $P_{\textrm{flip}}$ \\ \hline
	H$_{\textrm{ortho}}\to$H$_{\textrm{ortho}}$ & 0.659(1) \\
	H$_{\textrm{ortho}}\to$H$_{\textrm{para}}$ & 1.0 \\
	H$_{\textrm{para}}\to$H$_{\textrm{para}}$ & 0.0 \\
	H$_{\textrm{para}}\to$H$_{\textrm{ortho}}$ & 1.0 \\ \hline
    \end{tabular}
\caption[]{Molecular hydrogen scattering interactions with spin-flip probabilities as implemented in our MCNPX model using ENDF/B-VII cross sections~\cite{Chadwick2011}. The uncertainty for  H$_{\textrm{ortho}}\to$H$_{\textrm{ortho}}$ is due to the neutron energy spectrum.}
\label{tab:spinflip_reactions_H}
\end{center}
\end{table}
There are four possibilities of initial and final states for neutron scattering from a hydrogen molecule (see table \ref{tab:spinflip_reactions_H}). H$_{\textrm{para}}\to$H$_{\textrm{para}}$ scattering events do not change the spin of the neutron for the neutron energies relevant to the NPDGamma experiment. All H$_{\textrm{para}}\to$H$_{\textrm{ortho}}$ and H$_{\textrm{ortho}}\to$H$_{\textrm{para}}$ scattering events are inelastic in nature and necessarily cause a spin flip. A neutron with energy well below 14.7~meV does not have sufficient energy to undergo an inelastic spin-flip scatter from a parahydrogen molecule such that all interactions with parahydrogen do not change the neutron spin, unless the neutron has been upscattered above the 14.7~meV threshold. An H$_{\textrm{ortho}}\to$H$_{\textrm{ortho}}$ scattering event may cause a spin-flip if the molecular spin is orthogonal to the neutron spin, which follows the spin flip probability in equation \ref{eq:spin_flip_prob}. The Young-Koppel gas model as well as the Keinert-Sax liquid model were used separately in order to estimate the contributions of these scattering processes to the depolarization of neutrons for the NPDGamma experiment.

\subsubsection{Young-Koppel free gas model treatment}
Young and Koppel~\cite{Young1964} give analytical expressions for the differential scattering cross sections for orthohydrogen and parahydrogen gases with the inclusion of spin correlations, vibrations, and rotations. The momentum transfer is given by $\vec{K} = \vec{k} - \vec{k}'$, and the recoil energy of the molecule is given by,
\begin{equation}
E_{\textrm{rec}} = \frac{m_{\textrm{n}}}{m_{\textrm{m}}} \big( E_{\textrm{n}} - 2 \sqrt{E_{\textrm{n}} E_{\textrm{n}}'} (\hat{k} \cdot \hat{k}') + E_{\textrm{n}}'\big) = \frac{K^2\hbar^2}{4 m_{\textrm{n}}},
\label{eq:recoil_energy}
\end{equation}
where $E_{\textrm{n}}$ and $E_{\textrm{n}}'$ are the initial and final neutron energies, $\hat{k}$ and $\hat{k}'$ are the initial and final neutron velocity vectors, $m_{\textrm{n}}$ is the neutron mass, and $m_{\textrm{m}}$ is the mass of the molecule. The energy balance for the scattering event is given by,
\begin{equation}
\Delta E_{\textrm{n}} = E_{\textrm{n}}+E_{\textrm{rot}}(j_{\textrm{i}})-(E_{\textrm{n}}'+E_{\textrm{vib}}(n)+E_{\textrm{rot}}(j_{\textrm{f}})+E_{\textrm{rec}}),
\end{equation}
where $E_{\textrm{vib}}(n)$ is the vibrational energy of the molecule in the $n$th mode. At low temperatures, the molecular vibrational states can be ignored because molecules are in the vibrational ground state and cold neutrons do not have sufficient energy to excite higher order states. The Y-K model ignores intermolecular interactions and the cross section is given by the molecular ``self''-term. The simplified cross sections (for $J=0, 1$ )for the spin isomer transitions in hydrogen can be expressed analytically as follows from Zoppi~\cite{Zoppi1993}\cite{Celli1999a},
\begin{equation}
\begin{array}{ll}
\frac{d^2\sigma}{d\Omega dE'}|_{o \to o} = & 2 C (a_{\textrm{coh}}^2+\frac{2}{3}a_{\textrm{inc}}^2) (J_0(\alpha)^2 + 2 J_2(\alpha)^2), \\
\frac{d^2\sigma}{d\Omega dE'}|_{o \to p} = & -6 C a_{\textrm{inc}}^2 J_1(\alpha)^2, \\
\frac{d^2\sigma}{d\Omega dE'}|_{p \to p} = & 2 C a_{\textrm{coh}}^2  J_0(\alpha)^2, \\
\frac{d^2\sigma}{d\Omega dE'}|_{p \to o} = & -\frac{2}{3} C a_{\textrm{inc}}^2 J_1(\alpha)^2, \\
\end{array}
  \label{eq:YKxs}
\end{equation}
where $C = \sqrt{\frac{E_{\textrm{n}}'}{\pi E_{\textrm{n}} E_{\textrm{rec}} k_B T}} \exp{\big[-\frac{(\Delta E_{\textrm{n}})^2}{4 E_{\textrm{rec}} k_B T}\big]}$ is a factor common to each, $a_{\textrm{coh}}$ and $a_{\textrm{inc}}$ are the bound coherent and incoherent nuclear scattering lengths, $J_l(\alpha)$ is the spherical Bessel function of order $l$, $d$ is the equilibrium separation between the atoms in the molecule, and $\alpha=\frac{1}{2}Kd$. The $3J$ symbol and integral terms have been evaluated in order to supply concise expressions for the $J=0, 1$ transitions, which are the only terms that contribute for cold neutrons and liquid hydrogen in the vicinity of 15~K.

For the Y-K gas model, the probability of a spin flip interaction for H$_{\textrm{para}}\to$H$_{\textrm{ortho}}$ and H$_{\textrm{ortho}}\to$H$_{\textrm{para}}$ transitions is unity. Cold neutrons do not have sufficient energy for spin flip interactions from H$_{\textrm{para}}\to$H$_{\textrm{para}}$ scattering events. The depolarization factor for H$_{\textrm{ortho}}\to$H$_{\textrm{ortho}}$ transitions and 5~meV neutrons is determined from the cross section~\cite{Zoppi1993} in equation \ref{eq:YKxs} and is given by
\begin{equation}
R = \frac{\sigma_{\textrm{coh}} -\frac{2}{9}\sigma_{\textrm{incoh}}}{\sigma_{\textrm{coh}} +\frac{2}{3}\sigma_{\textrm{incoh}}} = -0.29.
\label{eq:YKR}
\end{equation}
The Y-K model only includes hydrogen transitions and does not incorporate spin-flip scattering from apparatus components.

\subsubsection{Keinert-Sax kernel model treatment}
The basis for hydrogen scattering kernels is the Young and Koppel model, which accurately describes rotational, vibrational, spin correlations, and free translations for gaseous hydrogen. However, translational modes in liquid hydrogen are not free and the Young and Koppel model is not sufficient to describe the scattering of cold neutrons from liquid hydrogen. The Keinert and Sax model~\cite{Keinert1987} improves the description of the translational modes compared to the Young and Koppel model in order to more accurately describe scattering from liquid hydrogen.
The scattering kernels compile for MCNPX from MacFarlane~\cite{Macfarlane} are based on the hindered translation model from Keinert and Sax and also incorporate interference between scattered waves from different molecules. The hydrogen molecules are represented as if there are solid-like clusters of approximately 20 molecules that diffuse through the liquid. The MacFarlane model better describes the steep drop in the parahydrogen cross section above 20~meV due to intermolecular spin correlation and the drop around 3-4~meV due to intramolecular interference~\cite{Macfarlane} when compared to the Young and Koppel gas model.
 
Despite the fact that MCNPX does not track spin or determine when a spin exchange scattering event takes place, the kinematics of a neutron interacting with orthohydrogen or parahydrogen carries the signature of spin exchange and this is used to determine whether a spin-flip interaction has taken place. The first excited state of the hydrogen molecule with $J=1$ is 14.7~meV above the ground state. Therefore, the signature of the H$_{\textrm{ortho}}\to$H$_{\textrm{para}}$ transition is a substantial increase in the energy of the neutron with an expected value near 14.7~meV, while the H$_{\textrm{para}}\to$H$_{\textrm{ortho}}$ transition signature is an energy decrease of the same magnitude.
\begin{figure}[h!]
	\centering
	\includegraphics{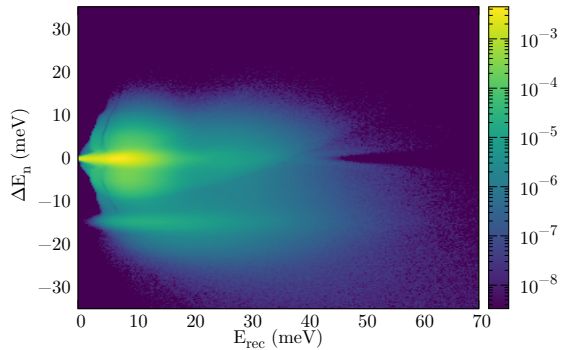}
	\caption{Parahydrogen scattering kinematics from MCNPX in $\Delta E_{\textrm{n}}$ and $E_{\textrm{rec}}$ (eq. \ref{eq:recoil_energy}) space (color online). The color scale indicates probability of an interaction occurring in each ($E_{\textrm{rec}}$, $\Delta E_{\textrm{n}}$) bin. The H$_{\textrm{para}}\to$H$_{\textrm{ortho}}$ to transition is the broad region in $E_{\textrm{rec}}$ at $\Delta E_{\textrm{n}} = -14.7$~meV}.
	\label{fig:para_recoil_scatter}
\end{figure}
\begin{figure}[h!]
	\centering
	\includegraphics{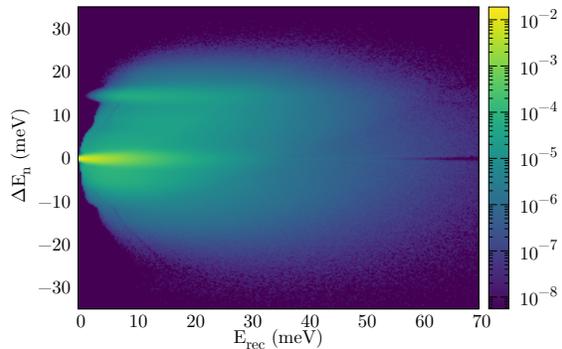}
	\caption{Orthohydrogen scattering kinematics from MCNPX in $\Delta E_{\textrm{n}}$ and $E_{\textrm{rec}}$ (eq. \ref{eq:recoil_energy}) space(color online). The color scale indicates probability of an interaction occurring in each ($E_{\textrm{rec}}$, $\Delta E_{\textrm{n}}$) bin.  The H$_{\textrm{ortho}}\to$H$_{\textrm{para}}$ to transition is the broad region in $E_{\textrm{rec}}$ at $\Delta E_{\textrm{n}} = 14.7$~meV}.
	\label{fig:ortho_recoil_scatter}
\end{figure}

A broad and flat input neutron source that does not match the FNPB energy spectrum was used to generate a map of the kinematics in $\Delta E_{\textrm{n}} = E_{\textrm{n}}' - E_{\textrm{n}}$ and $E_{\textrm{rec}} = \frac{K^2\hbar^2}{4 m_{\textrm{n}}}$ space for neutron scattering from parahydrogen and orthohydrogen. This was done for illustration and to determine whether the kinematics could be used as a marker for spin-flip scattering interactions by isolating a region around $\Delta E_{\textrm{n}} = \pm 15$~meV. Figures \ref{fig:para_recoil_scatter} and \ref{fig:ortho_recoil_scatter}, parahydrogen and orthohydrogen respectively, illustrate two dimensional histograms of the kinematics as calculated in MCNPX 2.7 using ENDF-VII~\citep{Chadwick2011} cross sections. 

The regions near $\pm$14.7~meV for each were fit to a 2 dimensional Gaussian in $\Delta E_{\textrm{n}}$ and $E_{\textrm{rec}}$. A 3$\sigma$ cutoff from these fits was used in the depolarization calculation such that any scattering event with $\Delta E_{\textrm{n}}$ and $E_{\textrm{rec}}$ inside 3$\sigma$ is labeled as a H$_{\textrm{ortho}}\to$H$_{\textrm{para}}$ or H$_{\textrm{para}}\to$H$_{\textrm{ortho}}$ and the neutron spin is flipped with unit probability. If the event is outside of this ellipse, the interacting species and table \ref{tab:spinflip_reactions_H} are used to  determine whether the interaction is H$_{\textrm{para}}\to$H$_{\textrm{para}}$  or H$_{\textrm{ortho}}\to$H$_{\textrm{ortho}}$.

\section{Monte Carlo calculations of spin-flip scattering}
Two separate Monte Carlo models have been developed in order to estimate the depolarization correction for the NPDGamma experiment. The first model uses the Y-K gas cross sections in order to estimate the contribution from orthohydrogen and parahydrogen. For the second, MCNPX 2.7 was modified in order to provide neutron spin state tracking throughout particle histories such that the neutron spin could be flipped probabilistically and the signal contributions from each spin state recorded for each detector. 

\subsection{Gaseous cross-section based simulation}
While the Y-K model does not adequately describe the translational modes in the liquid phase~\cite{Elliott1967}, the spin-correlation assumptions in the Y-K model are valid. The Y-K cross sections in equation \ref{eq:YKxs} are used to directly model the average polarization of a neutron beam as it passes through a parahydrogen/orthohydrogen gaseous mixture. 

The source neutron beam in this model is initialized with 100\% polarization and is monochromatic with an energy of 5~meV. The mean-free-path of the neutron is calculated and the neutron is propagated to a new location inside the hydrogen volume and the event type is chosen probabilistically at that location according to the value of $R$ in equation \ref{eq:YKR}. If the event is a capture event, the neutron history is terminated, the average polarization is recorded, and a new neutron track is generated at the source. If the event is a scattering event, then the neutron spin is flipped as appropriate, the kinematics of the collision are determined, and the calculation continues with a new determination of the mean-free-path and calculation continues until the neutron track is terminated when it has either escaped the liquid hydrogen volume or has captured on hydrogen.

\subsection{MCNPX simulation}
MCNPX does not have any internal way of defining and tracking neutron spin. This capability was implemented by adding a spin state tag for neutrons. 
As neutrons propagate in MCNPX, all information regarding the scattering material, interaction type, outgoing momentum, and possible newly created particles is determined internally. An additional step was incorporated to probabilistically change the neutron spin tag from spin up to spin down at each scattering event. Hydrogen in plastics and rubber components in other sections of the apparatus is denoted as $H_{\textrm{bound}}$.
\begin{table}
\begin{center}
    \begin{tabular}{| l | c |}
    \hline
    Isotope & $P_{\textrm{flip}}$ \\ \hline
	H$_{\textrm{bound}}$ & 0.659(1)\\
	$^{6}$Li & 0.0002(1) \\
	$^{7}$Li & 0.43(5) \\
	$^{14}$N & 0.022(3) \\
	$^{27}$Al & 0.0033(3) \\ 
	$^{35}$Cl & 0.0326(7) \\
	$^{37}$Cl & 0.00034(4) \\
	$^{55}$Mn & 0.012(5) \\
	$^{63}$Cu & 0.00041(9) \\
	$^{65}$Cu & 0.016(1) \\
	Zn & 0.0011(5) \\ \hline
    \end{tabular}
\caption{Isotopes with spin flip probabilities determined using Sears cross sections~\cite{Sears1992} as implemented in the model along with uncertainties due to averaging over the neutron energy spectrum. Zinc is applied over all natural isotopes rather than specific isotopes.}
\label{tab:spinflip_reactions}
\end{center}
\end{table}

The MCNPX calculation incorporates implicit capture, such that capture events produce secondary particles and reduce the weight of that neutron and continue propagating it. The materials in the MCNPX model of the NPDGamma experiment contain complicated alloys rather than mono-isotopic materials, and MCNPX handles the bookkeeping for the selection of the isotope involved in a scattering event. The isotopes relevant to the NPDGamma experiment with non-zero nuclear spin and an incoherent cross section from Sears~\cite{Sears1992} are shown in table \ref{tab:spinflip_reactions}. In the case of H$_{\textrm{bound}}$, the spin flip probability corresponds to the probability of a non-absorption event multiplied by a factor of $\frac{2}{3}$ due to the lack of separate coherent and incoherent cross sections in MCNPX. The material definitions in the simulation used ENDF/B-VII cross sections~\cite{Chadwick2011} for the MCNPX calculations.

\begin{figure}[h!]
	\centering
	\includegraphics{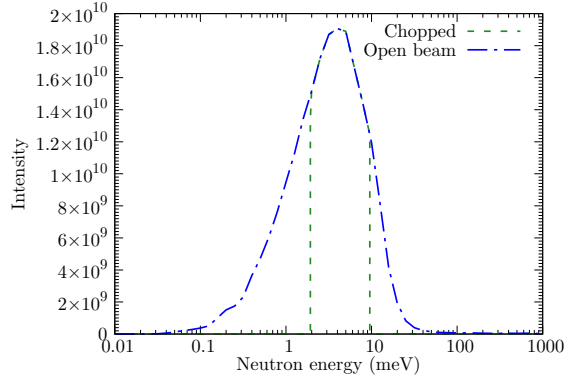}
	\caption{Input neutron spectrum used in the MCNPX calculations with approximate cutoffs for a typical chopped and open beam spectra for the NPDGamma experiment~\cite{Fomin2015}.}.
	\label{fig:fnpbchopped}
\end{figure}

The source neutron beam begins with 100\% polarization and has the energy spectrum shown in figure \ref{fig:fnpbchopped}. This spectrum is representative of the spectrum determined by the beamline chopper settings~\cite{Fomin2015} without frame overlap. As neutrons propagate in the model and create $\gamma$-rays, the $\gamma$-rays inherit the tag indicating the spin of the neutron as the $\gamma$-ray was created. Additionally, Compton electrons inherit this same tag as they are created through $\gamma$-ray scattering processes. 

The average polarization is determined by binning the energy deposition in each detector by the spin state tag, and a user supplied subroutine performs these tallies. The geometrically averaged polarization as measured for detector, $l=0...47$, is given by,
\begin{equation}
P^l_{\textrm{ave}}=\frac{E^l_{\uparrow}-E^l_{\downarrow}}{E^l_{\uparrow}+E^l_{\downarrow}},
\label{eq:p_ave}
\end{equation}
where $E^l_{\uparrow}$ and $E^l_{\downarrow}$ are the spin up and spin down energy deposition tallies respectively. The detector rings, $r$, are numbered from 1 to 4 with ring 1 being the most upstream ring. Each detector ring should observe approximately the same average polarization and downstream rings should observe a slightly lower average polarization because further downstream detector rings measure signals from neutrons that have penetrated deeper into the liquid and are therefore more likely to have undergone a spin-flip scattering event. Additionally, the penetration depth for neutrons is dependent on the neutron energy, and thus this method of determining the polarization correction captures both the geometric ring-dependence and energy dependence of $\Delta^{r}_{\textrm{dep}}(\lambda)$.

\section{Calculated polarization correction}
The MCNPX model provides the versatility to calculate the $\Delta^{r}_{\textrm{dep}}$ correction for multiple target materials in addition to hydrogen. In particular, this correction factor was calculated for auxiliary NPDGamma targets such as the chlorine-35 target and the aluminum 6061 target.

\subsection{Chlorine-35}
The chlorine-35 parity-violating asymmetry is large and well-known, with a world average of $(2.39 \pm 0.136) \times 10^{-5}$~\cite{Avenier1985}\cite{Vesna1982}\cite{Mitchell2004}. A Teflon target filled with carbon tetrachloride was used as a systematic check on the apparatus components. Natural chlorine is 75.8\% chlorine-35 and 24.2\% chlorine-37\cite{Meija2016}, and chlorine-35 has a significant spin flip probability (see table \ref{tab:spinflip_reactions}). The polarization correction for the chlorine target is $0.9840\pm0.0003$. Since the chlorine target is thin rather than comparable to the length of the detector arrays, there is no ring dependence to the polarization correction for chlorine.

\subsection{Aluminum-27}
\begin{figure}[h]
  \centering
\includegraphics{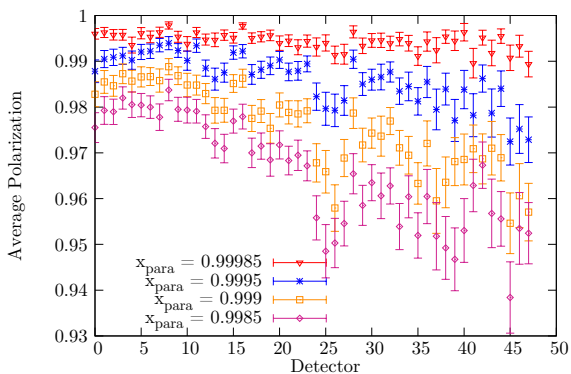}
\caption{$\Delta^{r}_{\textrm{dep}}$ for neutrons that capture on aluminum 6061 containing cells as a function of parahydrogen concentration and shown for each detector.}
  \label{fig:hydro_ALAUX_vessel_conc}
\end{figure}

The apparatus components, including the hydrogen target vessel and vacuum windows, are constructed of aluminum 6061 and neutron capture on these components can carry a parity violating signal that is a dilution factor in the measurement of the hydrogen asymmetry. The polarization correction for the apparatus aluminum is calculated using the MCNPX model and is dependent on the orthohydrogen concentration in the hydrogen vessel. The results are shown in figure \ref{fig:hydro_ALAUX_vessel_conc} and tabulated for each ring in table \ref{tab:hydrogen_ALAUX}.

There is a detector ring dependence seen in \ref{fig:hydro_ALAUX_vessel_conc} that is the result of the geometry of the target vessel and the fact that it is distributed over $\approx$ 30~cm. Downstream detectors are more likely than upstream detectors to detect $\gamma$-rays from neutrons that travel further through the liquid hydrogen volume before being captured. Neutrons that travel further through the liquid hydrogen are more likely to have scattered before being captured and therefore are more likely to have experienced a spin-flip scattering event than neutrons that captured a short distance into the target.
\begin{table}
\begin{center}
\label{tab:hydrogen_ALAUX}
\begin{tabular}{| c | c c |}
\hline
Ring & $\Delta^{r}_{\textrm{dep}}(x_{\textrm{o,lower}})$ & $\Delta^{r}_{\textrm{dep}}(x_{\textrm{o,upper}})$ \\ \hline
1 & 0.9961(3) & 0.9799(8) \\
2 & 0.9955(3) & 0.9722(8) \\
3 & 0.9939(5) & 0.9581(13) \\
4 & 0.9941(6) & 0.9556(19) \\
\hline
Total & 0.9953(2) & 0.9724(5) \\
\hline
\end{tabular}
\caption[Table caption text]{Polarization correction for the apparatus aluminum at the lower and upper limit orthohydrogen concentrations for each ring and averaged over the detector array.}
\end{center}
\end{table}

\subsection{Hydrogen}
Similar to the aluminum calculation, the $\Delta^{r}_{\textrm{dep}}$ correction for hydrogen has a detector ring dependence due to being a distributed volume. $\Delta^{r}_{\textrm{dep}}$ is sensitive to the orthohydrogen concentration ($x_{\textrm{o}}$) and calculations were performed at the upper and lower bound orthohydrogen concentrations that were determined along with the measurement of the parahydrogen scattering cross-section~\cite{Grammer2015}, which are 0.15\% and 0.015\%, respectively. Since there was no in situ method of determining the $x_{\textrm{o}}$ for the NPDGamma experiment, the $x_{\textrm{o}}$ distribution is assumed to be flat between the upper and lower bounds, with an uncertainty given by $\frac{1}{\sqrt{12}}(x_{\textrm{o,lower}} - x_{\textrm{o,upper}})$. Similarly, the polarization correction is given by
\begin{equation}
\Delta^{r}_{\textrm{dep,ave}} = \frac{1}{2}\left[\Delta^{r}_{\textrm{dep}}(x_{\textrm{o,lower}}) + \Delta^{r}_{\textrm{dep}}(x_{\textrm{o,upper}})\right],
\end{equation}
with uncertainty equal to 
\begin{equation}
\sigma^{r}_{\textrm{dep,ave}} = \sqrt{\frac{1}{12}}\left[\Delta^{r}_{\textrm{dep}}(x_{\textrm{o,lower}}) - \Delta^{r}_{\textrm{dep}}(x_{\textrm{o,upper}})\right].
\end{equation}

\begin{figure}[!h]
	\centering
	\includegraphics{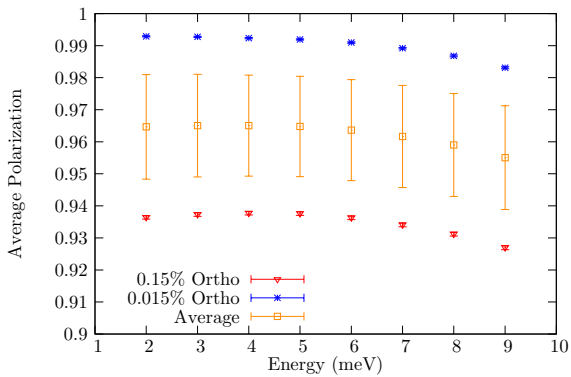}
	\caption{Polarization correction averaged over the detector rings as a function of initial neutron energy.}
	\label{fig:energy_scan}
\end{figure}

The neutron energy dependence of $\Delta^{r}_{\textrm{dep}}(\lambda)$ was calculated using a series of calculations with monochromatic neutron source terms and varying the energy. The results of this calculation are shown in figure \ref{fig:energy_scan}, where the results from the four rings have been averaged together for each neutron energy. The small deviation from 100\% polarization at zero orthohydrogen concentration is due to the upscattering of neutrons above the spin-flip threshold as well as the combined kinetic energy of the molecules and neutron above the spin-flip threshold.

\begin{figure}[!h]
  \centering
  \includegraphics{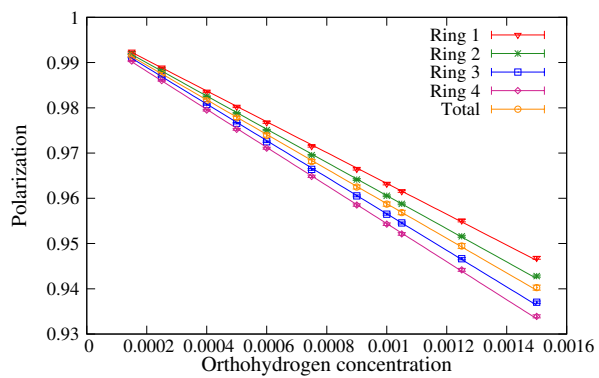}
  \caption{Polarization correction for each ring as a function of orthohydrogen concentration.}
  \label{fig:depol_scan}
\end{figure}

Simulations with the MCNPX model were performed at multiple orthohydrogen concentrations within this range and the results are shown in figure \ref{fig:depol_scan}. The polarization upon capture is a linear function of orthohydrogen concentration within this range. Calculations with monochromatic neutron beams were also performed and show that the polarization correction decreases with increasing energy within the typical NPDGamma neutron spectrum range, which is below the 14.7~meV rotational transition energy. The average contribution to $\Delta^{r}_{\textrm{dep}}$ for hydrogen due to materials other than orthohydrogen and parahydrogen in table \ref{tab:spinflip_reactions} is $2\times10^{-4}$ and is independent of orthohydrogen concentration, since the majority of this contribution is due to spin-flip scattering that takes place upstream of the target vessel in vacuum windows and other apparatus components.
\begin{table}[!h]
\begin{center}
    \begin{tabular}{| l | c | c |}
    \hline
    Ring & $\Delta^{r}_{\textrm{dep}}(x_{\textrm{o,lower}})$ & $\Delta^{r}_{\textrm{dep}}(x_{\textrm{o,upper}})$ \\ \hline
	1 & $0.99275(11)$ & $0.94712(28)$ \\
	2 & $0.99233(9)$ & $0.94304(24)$ \\
	3 & $0.99144(10)$ & $0.93735(25)$ \\
	4 & $0.99083(13)$ & $0.93417(33)$ \\ \hline
	Total & $0.99187(21)$ & $0.94053(56)$ \\ \hline
    \end{tabular}
\caption[Table caption text]{Ring dependent polarization correction for liquid hydrogen at the lower and upper limit orthohydrogen concentrations for each ring and averaged over the detector array.}
\label{tab:min_orthoa}
\end{center}
\end{table}

The uncertainty due to the unknown orthohydrogen concentration is much larger than both the uncertainty due to the energy dependence and the ring dependence, thus the polarization correction is averaged over the energy spectrum and over all detector rings to produce a single value that applies to all detectors. The resulting polarization correction from the MCNPX calculation is between 0.99187 and 0.94053 with an average of $0.966 \pm 0.015$. By comparison, the Y-K model uses a monochromatic neutron beam at 5~meV and yields an average polarization between $0.97$ and $0.9959$ with an average of $0.983 \pm 0.007$. In order to be conservative, we take the maximum and minimum of the results from the Y-K and MCNPX results and assign the bounds on $\Delta_{\textrm{dep}}$ to be between $0.94053$ and $0.9959$ and take the probability distribution between these to be flat. Therefore, the multiplicative polarization correction to the NPDGamma experiment, $\Delta_{\textrm{dep}}$, is determined to be $0.968 \pm 0.016$. This uncertainty is comparable to that of other multiplicative corrections for the NPDGamma experiment, namely the geometrical factors calculation~\cite{Grammer2018gf} and the measured beam polarization~\cite{Musgrave2018} and significantly smaller than the statistical uncertainty in the measurement.

\section{Acknowledgments}
This material is based upon work supported by the U.S. Department of Energy, Office of Science, Office of Basic Energy Sciences under contract number DE-AC05-00OR22725. This work was also supported by  DE-FG02-03ER41258.

\bibliography{papers_writeups-depol}

\begin{thebibliography}{10}
\expandafter\ifx\csname url\endcsname\relax
  \def\url#1{\texttt{#1}}\fi
\expandafter\ifx\csname urlprefix\endcsname\relax\def\urlprefix{URL }\fi
\expandafter\ifx\csname href\endcsname\relax
  \def\href#1#2{#2} \def\path#1{#1}\fi

\bibitem{Blyth2018}
D.~Blyth, J.~Fry, N.~Fomin, R.~Alarcon, L.~Alonzi, E.~Askanazi, S.~Bae{\ss}ler,
  S.~Balascuta, L.~Barr{\'{o}}n-Palos, A.~Barzilov, J.~D. Bowman, N.~Birge,
  J.~R. Calarco, T.~E. Chupp, V.~Cianciolo, C.~E. Coppola, C.~B. Crawford,
  K.~Craycraft, D.~Evans, C.~Fieseler, E.~Frle{\v{z}}, I.~Garishvili, M.~T.~W.
  Gericke, R.~C. Gillis, K.~B. Grammer, G.~L. Greene, J.~Hall, J.~Hamblen,
  C.~Hayes, E.~B. Iverson, M.~L. Kabir, M.~Maldonado-Vel{\'{a}}zquez,
  Y.~Masuda, J.~Mei, R.~Milburn, P.~E. Mueller, M.~Musgrave, H.~Nann,
  I.~Novikov, D.~Parsons, S.~I. Penttila, D.~Pocanic, A.~Ramirez-Morales,
  M.~Root, A.~Salas-Bacci, S.~Santra, S.~Schr{\"{o}}der, E.~Scott, P.-N. Seo,
  E.~I. Sharapov, F.~Simmons, W.~M. Snow, A.~Sprow, J.~Stewart, E.~Tang,
  Z.~Tang, X.~Tong, D.~J. Turkoglu, R.~Whitehead, W.~S. Wilburn, {First
  Observation of P-odd $\gamma$ Asymmetry in Polarized Neutron Capture on
  Hydrogen}, Physical Review Letters 13~(24) (2018) 14.

\bibitem{Gericke2011}
M.~T. Gericke, R.~Alarcon, S.~Balascuta, L.~Barr{\'{o}}n-Palos, C.~Blessinger,
  J.~D. Bowman, R.~D. Carlini, W.~Chen, T.~E. Chupp, C.~Crawford, S.~Covrig,
  M.~Dabaghyan, N.~Fomin, S.~J. Freedman, T.~R. Gentile, R.~C. Gillis, G.~L.
  Greene, F.~W. Hersman, T.~Ino, G.~L. Jones, B.~Lauss, M.~Leuschner, W.~R.
  Lozowski, R.~Mahurin, Y.~Masuda, J.~Mei, G.~S. Mitchell, S.~Muto, H.~Nann,
  S.~a. Page, S.~I. Penttil{\"{a}}, W.~D. Ramsay, A.~Salas-Bacci, S.~Santra,
  M.~Sharma, P.-N. Seo, E.~I. Sharapov, T.~B. Smith, W.~M. Snow, W.~S. Wilburn,
  V.~Yuan, {Measurement of parity-violating gamma-ray asymmetry in the capture
  of polarized cold neutrons on protons}, Physical Review C 83~(1) (2011)
  15505.

\bibitem{Fomin2015}
N.~Fomin, G.~L. Greene, R.~R. Allen, V.~Cianciolo, C.~Crawford, T.~M. Tito,
  P.~R. Huffman, E.~B. Iverson, R.~Mahurin, W.~M. Snow, {Fundamental neutron
  physics beamline at the spallation neutron source at ORNL}, Nucl. Instrum.
  Methods Phys. Res., Sect. A 773 (2015) 45--51.

\bibitem{Balascuta2012}
S.~Balascuta, R.~Alarcon, S.~Bae{\ss}ler, G.~Greene, A.~Mietke, C.~Crawford,
  R.~Milburn, S.~Penttila, J.~Prince, J.~Sch{\"{a}}dler, {The implementation of
  a super mirror polarizer at the SNS fundamental neutron physics beamline},
  Nuclear Instruments and Methods in Physics Research, Section A: Accelerators,
  Spectrometers, Detectors and Associated Equipment 671 (2012) 137--143.

\bibitem{Gericke2006}
M.~T. Gericke, J.~D. Bowman, R.~D. Carlini, T.~E. Chupp, K.~P. Coulter,
  M.~Dabaghyan, M.~Dawkins, D.~Desai, S.~J. Freedman, T.~R. Gentile, R.~C.
  Gillis, G.~L. Greene, F.~W. Hersman, T.~Ino, G.~L. Jones, M.~Kandes,
  B.~Lauss, M.~Leuschner, W.~R. Lozowski, R.~Mahurin, M.~Mason, Y.~Masuda,
  G.~S. Mitchell, S.~Muto, H.~Nann, S.~A. Page, S.~I. Penttil{\"{a}}, W.~D.
  Ramsay, S.~Santra, P.-N. Seo, E.~I. Sharapov, T.~B. Smith, W.~M. Snow, W.~S.
  Wilburn, V.~Yuan, H.~Zhu, {Upper bounds on parity-violating gamma-ray
  asymmetries in compound nuclei from polarized cold neutron capture}, Physical
  Review C 74~(6) (2006) 065503.

\bibitem{Fry2017}
J.~Fry, R.~Alarcon, R.~Allen, E.~Askanazi, S.~Balascuta, L.~Barron-Palos,
  S.~Bae{\ss}ler, A.~Barzilov, C.~Blessinger, D.~Blyth, J.~D. Bowman, J.~R.
  Calarco, T.~E. Chupp, C.~E. Coppola, C.~Crawford, K.~Craycraft, M.~Dabaghyan,
  D.~Evans, J.~Favela, C.~Fieseler, N.~Fomin, W.~Fox, S.~Freedman,
  E.~Frle{\v{z}}, C.~Fu, C.~Garcia, I.~Garishvili, M.~T. Gericke, R.~C. Gillis,
  K.~Grammer, G.~L. Greene, J.~Hamblen, C.~Hayes, F.~W. Hersman, T.~Ino, E.~B.
  Iverson, G.~L. Jones, L.~Kabir, S.~Kucucker, B.~Lauss, Y.~Li, R.~Mahurin,
  M.~Maldonado-Velazquez, M.~McCrea, Y.~Masuda, J.~Mei, R.~Milburn, G.~S.
  Mitchell, P.~Mueller, S.~Muto, M.~Musgrave, H.~Nann, I.~Novikov, S.~Page,
  D.~Parsons, D.~Po{\v{c}}ani{\'{c}}, S.~I. Penttil{\"{a}}, W.~D. Ramsay,
  A.~Salas-Bacci, S.~Santra, P.-N.~N. Seo, E.~Sharapov, M.~Sharma, F.~Simmons,
  T.~Smith, W.~M. Snow, J.~Stuart, E.~Tang, Z.~Tang, J.~Thomison, T.~Tong,
  J.~Vanderwerp, S.~Waldecker, W.~S. Wilburn, W.~Xu, V.~Yuan, Y.~Zhang, {Status
  of the NPDGamma experiment}, Hyperfine Interactions 238~(1).

\bibitem{Seo2008}
P.-N. Seo, L.~Barr{\'{o}}n-Palos, J.~D. Bowman, T.~E. Chupp, C.~Crawford,
  M.~Dabaghyan, M.~Dawkins, S.~J. Freedman, T.~Gentile, M.~T. Gericke, R.~C.
  Gillis, G.~L. Greene, F.~W. Hersman, G.~L. Jones, M.~Kandes, S.~Lamoreaux,
  B.~Lauss, M.~B. Leuschner, R.~Mahurin, M.~Mason, J.~Mei, G.~S. Mitchell,
  H.~Nann, S.~A. Page, S.~I. Penttil{\"{a}}, W.~D. Ramsay, A.~Salas-Bacci,
  S.~Santra, M.~Sharma, T.~B. Smith, W.~M. Snow, W.~S. Wilburn, H.~Zhu,
  {High-efficiency resonant rf spin rotator with broad phase space acceptance
  for pulsed polarized cold neutron beams}, Physical Review Special Topics -
  Accelerators and Beams 11~(8) (2008) 84701.

\bibitem{Santra2010}
S.~Santra, L.~{Barrn Palos}, C.~Blessinger, J.~D. Bowman, T.~E. Chupp,
  S.~Covrig, C.~Crawford, M.~Dabaghyan, J.~Dadras, M.~Dawkins, M.~T. Gericke,
  W.~Fox, R.~C. Gillis, M.~B. Leuschner, B.~Lozowski, R.~Mahurin, M.~Mason,
  J.~Mei, H.~Nann, S.~I. Penttila, A.~Salas-Bacci, M.~Sharma, W.~M. Snow, W.~S.
  Wilburn, {A liquid parahydrogen target for the measurement of a
  parity-violating gamma asymmetry in n→+p→d+$\gamma$}, Nuclear Instruments
  and Methods in Physics Research, Section A: Accelerators, Spectrometers,
  Detectors and Associated Equipment 620~(2-3) (2010) 421--436.

\bibitem{Gericke2005a}
M.~T. Gericke, C.~Blessinger, J.~D. Bowman, R.~C. Gillis, J.~Hartfield, T.~Ino,
  M.~Leuschner, Y.~Masuda, G.~S. Mitchell, S.~Muto, H.~Nann, S.~A. Page, S.~I.
  Penttil{\"{a}}, W.~D. Ramsay, P.-N. Seo, W.~M. Snow, J.~Tasson, W.~S.
  Wilburn, {A current mode detector array for $\gamma$ ray asymmetry
  measurements}, Nuclear Instruments and Methods in Physics Research, Section
  A: Accelerators, Spectrometers, Detectors and Associated Equipment 540~(2-3)
  (2005) 328--347.

\bibitem{Musgrave2018}
M.~Musgrave, S.~Bae{\ss}ler, S.~Balascuta, L.~Barr{\'{o}}n-Palos, D.~Blyth,
  J.~Bowman, T.~Chupp, V.~Cianciolo, C.~Crawford, K.~Craycraft, N.~Fomin,
  J.~Fry, M.~Gericke, R.~Gillis, K.~Grammer, G.~Greene, J.~Hamblen, C.~Hayes,
  P.~Huffman, C.~Jiang, S.~Kucuker, M.~McCrea, P.~Mueller, S.~Penttil{\"{a}},
  W.~Snow, E.~Tang, Z.~Tang, X.~Tong, W.~Wilburn, {Measurement of the absolute
  neutron beam polarization from a supermirror polarizer and the absolute
  efficiency of a neutron spin rotator for the NPDGamma experiment using a
  polarized 3 He neutron spin-filter}, Nuclear Instruments and Methods in
  Physics Research Section A: Accelerators, Spectrometers, Detectors and
  Associated Equipment 895 (2018) 19--28.

\bibitem{Szymanski1994}
J.~J. Szymanski, J.~D. Bowman, P.~P.~J. Delheij, C.~M. Frankle, J.~Knudson,
  S.~Penttil{\"{a}}, S.~J. Seestrom, S.~H. Yoo, V.~W. Yuan, X.~Zhu, {Ion
  chamber system for neutron flux measurements}, Nucl. Instrum. Methods Phys.
  Res., Sect. A 340~(3) (1994) 564--571.

\bibitem{Dennison1927}
D.~M. Dennison, {A Note on the Specific Heat of the Hydrogen Molecule},
  Proceedings of the Royal Society A: Mathematical, Physical and Engineering
  Sciences 115~(771) (1927) 483--486.

\bibitem{Barron-Palos2011}
L.~Barr{\'{o}}n-Palos, R.~Alarcon, S.~Balascuta, C.~Blessinger, J.~D. Bowman,
  T.~E. Chupp, S.~Covrig, C.~B. Crawford, M.~Dabaghyan, J.~Dadras, M.~Dawkins,
  W.~Fox, M.~T. Gericke, R.~C. Gillis, B.~Lauss, M.~B. Leuschner, B.~Lozowski,
  R.~Mahurin, M.~Mason, J.~Mei, H.~Nann, S.~I. Penttil{\"{a}}, W.~D. Ramsay,
  A.~Salas-Bacci, S.~Santra, P.~N. Seo, M.~Sharma, T.~Smith, W.~M. Snow, W.~S.
  Wilburn, V.~Yuan, {Determination of the parahydrogen fraction in a liquid
  hydrogen target using energy-dependent slow neutron transmission}, Nuclear
  Instruments and Methods in Physics Research, Section A: Accelerators,
  Spectrometers, Detectors and Associated Equipment 659~(1) (2011) 579--586.

\bibitem{catalyst}
{Sigma-Aldridge Corp., 3050 Spruce Street, St. Louis, MO 63103, USA}, {Product
  Number 371254-250G.}

\bibitem{Chadwick2011}
M.~Chadwick, M.~Herman, P.~Oblo{\v{z}}insk{\'{y}}, M.~Dunn, Y.~Danon,
  A.~Kahler, D.~Smith, B.~Pritychenko, G.~Arbanas, R.~Arcilla, R.~Brewer,
  D.~Brown, R.~Capote, A.~Carlson, Y.~Cho, H.~Derrien, K.~Guber, G.~Hale,
  S.~Hoblit, S.~Holloway, T.~Johnson, T.~Kawano, B.~Kiedrowski, H.~Kim,
  S.~Kunieda, N.~Larson, L.~Leal, J.~Lestone, R.~Little, E.~McCutchan,
  R.~MacFarlane, M.~MacInnes, C.~Mattoon, R.~McKnight, S.~Mughabghab, G.~Nobre,
  G.~Palmiotti, A.~Palumbo, M.~Pigni, V.~Pronyaev, R.~Sayer, A.~Sonzogni,
  N.~Summers, P.~Talou, I.~Thompson, A.~Trkov, R.~Vogt, S.~van~der Marck,
  A.~Wallner, M.~White, D.~Wiarda, P.~Young, {ENDF/B-VII.1 Nuclear Data for
  Science and Technology: Cross Sections, Covariances, Fission Product Yields
  and Decay Data}, Nuclear Data Sheets 112~(12) (2011) 2887--2996.

\bibitem{mughabghab2006atlas}
S.~F. Mughabghab, {Atlas of Neutron Resonances}, 5th Edition, Elsevier, New
  York, 2006.

\bibitem{Grammer2015}
K.~B. Grammer, R.~Alarcon, L.~Barron-Palos, D.~Blyth, J.~D. Bowman, J.~Calarco,
  C.~Crawford, K.~Craycraft, D.~Evans, N.~Fomin, J.~Fry, M.~Gericke, R.~C.
  Gillis, G.~L. Greene, J.~Hamblen, C.~Hayes, S.~Kucuker, W.~S. Wilburn,
  {Measurement of the scattering cross section of slow neutrons on liquid
  parahydrogen from neutron transmission}, Phys. Rev. B 180301~(18) (2015)
  1--6.

\bibitem{Pelowitz2011}
D.~B. Pelowitz, {MCNPX User's Manual}, Los Alamos National Laboratory, 2011.

\bibitem{Young1964}
J.~Young, J.~Koppel, {Slow Neutron Scattering by Molecular Hydrogen and
  Deuterium}, Phys. Rev. 135~(3A) (1964) A603----A611.

\bibitem{Moon1969a}
R.~Moon, T.~Riste, W.~Koehler, {Polarization Analysis of Thermal-Neutron
  Scattering}, Phys. Rev. 181~(2) (1969) 920--931.

\bibitem{Zoppi1993}
M.~Zoppi, {Neutron scattering of homonuclear diatomic liquids}, Physica B:
  Condensed Matter 183~(3) (1993) 235--246.

\bibitem{Celli1999a}
M.~Celli, N.~Rhodes, A.~K. Soper, M.~Zoppi, {The total neutron cross section of
  liquid para-hydrogen}, J. Phys.: Condens. Matter 11~(50) (1999) 10229--10242.

\bibitem{Keinert1987}
J.~Keinert, J.~Sax, {Investigation of neutron scattering dynamics in liquid
  hydrogen and deuterium for cold neutron sources}, Kerntechnik 51 (1987) 19.

\bibitem{Macfarlane}
R.~E. MacFarlane, {Cold-moderator scattering kernel methods}, Tech. rep., Los
  Alamos National Laboratory (LANL), Los Alamos, NM (aug 1998).

\bibitem{Elliott1967}
R.~J. Elliott, W.~M. Hartmann, {Theory of neutron scattering from liquid and
  solid hydrogen}, Proceedings of the Physical Society 90~(3) (1967) 671--680.

\bibitem{Sears1992}
V.~F. Sears, {Neutron scattering lengths and cross sections}, Neutron News
  3~(3) (1992) 26--37.

\bibitem{Avenier1985}
M.~Avenier, G.~Bagieu, H.~Benkoula, J.~F. Cavaignac, A.~Idrissi, D.~H. Koang,
  B.~Vignon, R.~Wilson, {Parity non-conservation in the radiative capture of
  polarized neutrons by35Cl}, Nuclear Physics, Section A 436~(1) (1985) 83--92.

\bibitem{Vesna1982}
V.~Vesna, {\'{E}}.~Kolomenski, V.~Lobashev, V.~Nazarenko, A.~Pirozhkov,
  L.~Smotritski, Y.~Sobolev, N.~Titov, {Observation of parity nonconservation
  in the integrated $\gamma$ spectrum from (n, $\gamma$) reactions in Cl, Br,
  Cd, Sn, and La nuclei}, Soviet Journal of Experimental and Theoretical
  Physics Letters 36 (1982) 209.

\bibitem{Mitchell2004}
G.~S. Mitchell, C.~S. Blessinger, J.~D. Bowman, T.~E. Chupp, K.~P. Coulter,
  M.~Gericke, G.~L. Jones, M.~B. Leuschner, H.~Nann, S.~A. Page, S.~I.
  Penttil{\"{a}}, T.~B. Smith, W.~M. Snow, W.~S. Wilburn, {A measurement of
  parity-violating gamma-ray asymmetries in polarized cold neutron capture on
  35Cl, 113Cd, and 139La}, Nuclear Instruments and Methods in Physics Research
  Section A: Accelerators, Spectrometers, Detectors and Associated Equipment
  521~(2-3) (2004) 468--479.

\bibitem{Meija2016}
J.~Meija, T.~B. Coplen, M.~Berglund, W.~A. Brand, P.~{De Bi{\`{e}}vre},
  M.~Gr{\"{o}}ning, N.~E. Holden, J.~Irrgeher, R.~D. Loss, T.~Walczyk,
  T.~Prohaska, {Isotopic compositions of the elements 2013 (IUPAC Technical
  Report)}, Pure and Applied Chemistry 88~(3) (2016) 293--306.

\bibitem{Grammer2018gf}
K.~Grammer, D.~Blyth, J.~Bowman, N.~Fomin, G.~Greene, M.~Musgrave, E.~Tang,
  Z.~Tang, {Monte Carlo calculation and verification of the geometrical factors
  for the NPDGamma experiment}, Nuclear Instruments and Methods in Physics
  Research Section A: Accelerators, Spectrometers, Detectors and Associated
  Equipment 903 (2018) 302--308.

\end{thebibliography}

\end{document}